\documentstyle[prd,aps,epsf,eqsecnum]{revtex}
\begin{document}
\draft
\title{Effective equation of state for a spherically 
expanding pion plasma}
\author{Melissa A. Lampert 
\thanks{electronic mail: melissa.lampert@unh.edu}}
\address{Department of Physics,
University of New Hampshire,
Durham, NH 03824 \\
}
\author{Carmen Molina-Par{\'{\i}}s
\thanks{electronic mail: carmen@t6-serv.lanl.gov}}
\address{Theoretical Division,
Los Alamos National Laboratory,
Los Alamos, NM 87545 \\
}
\preprint{LA-UR-97-2058}
\date{\today}
\maketitle

\begin{abstract}
{Following a relativistic heavy ion collision, the quark-gluon plasma
produced eventually undergoes a chiral phase transition.  We assume
that during this phase transition one can describe the dynamics of the
system by the linear $\sigma$ model and that the expansion can be
thought of as mostly radial.  Because the $\sigma$ model is an
effective field theory there is an actual momentum cutoff (Landau
pole) in the theory at around 1 GeV.  Thus it is necessary to find
ways of obtaining a covariantly conserved, renormalized
energy-momentum tensor when there is a cutoff present (which breaks
covariance), in order to identify the effective equation of state of
this time evolving system. We show how to solve this technical problem
and then determine the energy density and pressure of the system as a
function of the proper time.  We consider different initial conditions
and search for instabilities which can lead to the formation of
disoriented chiral condensates (DCCs).  We find that the energy
density and pressure both decrease quickly, as is appropriate for a
rapidly cooling system, and that the energy is numerically conserved.}
\end{abstract}
%
\section{Introduction}
During a relativistic heavy ion collision, it is possible to create a
quark-gluon plasma, which will then cool and expand, leading to
hadronization.  At around the same time as hadronization, the system
also undergoes a chiral phase transition which breaks the SU(2)
$\times$ SU(2) symmetry and which then gives physical masses to the
pion and other particles. The out of equilibrium dynamics of this
phase transition can be very interesting if the expansion is fast
enough so that the phase transition resembles a quench. In this case
the effective pion mass can go negative for short periods of proper
time which leads to a temporary exponential growth of low momentum
domains where the isospin can point in a particular direction
(DCC). This is a topic of current interest \cite{ref:dcc}, since DCCs
may provide a signature of the chiral phase transition.  In an earlier
paper \cite{ref:mypaper} the time evolution of this chiral phase
transition, assuming a uniform spherically symmetric expansion into
the vacuum, was considered. The O(4) linear $\sigma$ model in the
large-$N$ approximation, which incorporates both nonequilibrium and
quantum effects, was studied. A range of initial conditions was
analyzed to determine which initial states could lead to the formation
of DCCs.  This radial expansion is interesting because it is expected
at late times in the plasma evolution, and also maximizes the cooling
rate and therefore the possibility of the phase transition resembling
a quench.

In this paper we go beyond previous work \cite{ref:mypaper} and study
the proper time evolutions of the effective hydrodynamic collective
variables of the system, namely the renormalized pressure and energy
density. We want to see to what extent previous intuition coming from
studying classical hydrodynamical models of particle production such
as Landau's model is confirmed.  This requires the solution of a new
technical issue, namely how one obtains a covariantly conserved
energy-momentum tensor when one has an actual cutoff in the theory.
As discussed in earlier work \cite{ref:cooperdcc,ref:phi4}, the linear
$\sigma$ model is an effective theory with a renormalized coupling
constant of order 10, in order to agree with low energy pion
properties. This leads to a maximum cutoff of around 1 GeV in order to
avoid problems at scales of the order of the Landau pole.  It has been
shown earlier \cite{ref:cooperdcc} that as long as the momentum cutoff
determined from the sum over mode functions is below the Landau pole,
there is a regime of cutoffs where the continuum renormalization group
flow is obeyed so that one can safely say that one is in the continuum
regime.  For this class of problems, renormalization methods based on
formal schemes such as dimensional regularization approaches are not
very useful.  Here we have a real cutoff in physical momentum as a
result of using an effective field theory, and we have to be very
careful in order to obtain the correct covariantly conserved
energy-momentum tensor which leads to our definitions of comoving
energy density and pressure variables.  The tools we use to correctly
renormalize the energy-momentum tensor are first to introduce the
physical cutoff $\Lambda$ and then to analyze the divergences in all
the components of the energy-momentum tensor using an adiabatic
expansion of the mode functions.  By comparing our results with a
covariant point-splitting approach \cite{ref:christensen}, we can
identify those terms which should survive in the covariant limit. This
technique allows us to obtain finite energy densities and pressures
which then enable us to study the effective equation of state for the
evolving plasma.

In this paper, we derive the energy-momentum tensor for the O(4)
linear $\sigma$ model, and show how to regularize and renormalize it
in order to obtain the physical energy density and pressure. We study
initial conditions which have instabilities that lead to DCC formation
and also stable initial configurations, and examine the energy density
and pressure in both situations. We also consider the conservation of
the energy-momentum tensor as a check on our numerical methods.

The paper is organized as follows. In Sec.~\ref{sec:tensor} we discuss
the model and coordinate system used, and show how to construct the
energy-momentum tensor. Then in Sec.~\ref{sec:renorm} we explicitly
describe the scheme used to renormalize this tensor in order to
perform a numerical simulation with a physical momentum cutoff.  In
Sec.~\ref{sec:results} we numerically examine the conservation of
energy and the equation of state of the system. Finally in
Sec.~\ref{sec:conc} we discuss the results and provide some concluding
remarks.
 
%
\section{Energy-Momentum Tensor}
\label{sec:tensor}
\subsection{Equations of Motion and Coordinate System}
\label{subsec:eom}
We consider the O(4) linear $\sigma$ model, with the classical action
given by \cite{ref:mypaper}
\begin{eqnarray}
   S[\Phi_i,\chi;j_i] = \int d^4 x \sqrt{-g(x)} \left(
      {\cal L}[\Phi_i, \chi] +  j_i \Phi_i \right)
\>,
\end{eqnarray}
where the mesons are in a O(4) vector representation
\begin{equation}
   \Phi_i = (\sigma, \vec \pi)
\>.
\end{equation}
The Lagrangian density is
\begin{eqnarray}
   {\cal L}[\Phi_i, \chi] =
      \frac{1}{2}
      g_{\mu\nu}(x) [ \partial^\mu \Phi_i(x) ][ \partial^\nu \Phi_i(x) ]
      - \frac{1}{2} \chi(x) \Phi_i^2(x) + \frac{v^2}{2} \chi(x) 
      + \frac{1}{4 \lambda} \chi^2(x)
\>.
\label{eq:lag}
\end{eqnarray}
In order to quantize the system, we construct the generating
functional of the connected Green's functions and carry out the path
integral to obtain the following quantum effective action (to leading
order in $N$) \cite{ref:mypaper}
\begin{equation}
   \Gamma[\phi_{i},\chi ; j_i]= S[\phi_{i},\chi ; j_i]
      + \frac{iN}{2} {\rm Tr} \ln G_0^{-1}(x,x'; \chi)
\>,
\end{equation}
where
\begin{equation}
   G_0^{-1}(x, x'; \chi) = \left[\Box + \chi(x) \right] 
      \delta^4(x,x') / \sqrt{-g(x)}
\>,
\end{equation}
and
\begin{eqnarray}
   \phi_{i}(x) &\equiv& \langle \Phi_{i}(x)\rangle,
   \nonumber \\ 
   \chi(x) &\equiv& \langle \chi(x)\rangle
\>.
\end{eqnarray}
By varying the effective action with respect to $\phi_i$ and $\chi$, we
obtain the following equations of motion for the mean field $\phi_i$
\begin{equation}
   \left[\Box + \chi(x)\right] \phi_i(x) = 
      j_i(x) \equiv H \delta_{i0}\
\>,
\label{eq:eom}
\end{equation}
and the constraint equation for the effective mass squared $\chi$
\begin{equation}
   \chi(x) = \lambda [-v^2 + \phi_i^2(x) - i N G_0(x,x; \chi)]
\>.
\label{eq:chieom}
\end{equation}
Notice that $G_0(x, x'; \chi)$ depends on $\chi$ through its definition,
and it fulfills the following dynamical equation 
\begin{equation}
  \left[ \Box + \chi(x) \right] G_0(x, x'; \chi)  = 
     \delta^4(x,x') / \sqrt{-g(x)}
\>.
\end{equation}
The current has only a non-vanishing component in the zero ($\sigma$)
direction in order to give mass to the pion \cite{ref:mypaper}.  There
are three parameters in the model: the mass of the pion $m = m_{\pi}$;
the value of $H = f_\pi m_\pi^2$, which gives the vacuum expectation
value $\langle \sigma \rangle = f_\pi$ (using PCAC); and the coupling
constant $\lambda$, which is determined by fitting to low energy
$\pi-\pi$ scattering data \cite{ref:mypaper,ref:cooperdcc}.

The picture one gets from hydrodynamical simulations of heavy ion
collisions is that the energy density is initially in a Lorentz
contracted disk which expands first in the longitudinal direction and
becomes three dimensional at late times.  At early times one expects
the velocity to scale approximately as $z/t$, where $z$ is the
longitudinal direction, since the effective longitudinal size goes to
zero with center of mass energy. This leads to the energy density
approximately becoming a function only of the longitudinal fluid
proper time variable $ \tau_z= (t^2-z^2)^{1/2}$.  At later times when the
expansion is more spherical and the initial distribution looks more
like a ``point'', one expects that the velocity scales as $v=r/t$ and
the energy density then becomes a function of the spherical fluid
variables:
\begin{eqnarray}
   \tau & = & {(t^2 - r^2)}^{1/2},
   \nonumber \\
   \eta & = & {\rm arctanh} (r/t) =
      \frac{1}{2} \ln \left( \frac{t + r}{t - r} \right)
\>,
\end{eqnarray}
where $t = \tau \cosh \eta$ and $r =\tau \sinh \eta$
\cite{ref:cooperlandau}. We restrict the range of these variables to
the future light cone, namely $0 \leq \tau < \infty$ and $0 \leq \eta
< \infty$. The coordinates $(\tau, \eta, \theta, \varphi)$ are useful
to describe a spherically symmetric expansion of a plasma when one is
in a hydrodynamical scaling regime. A spherical expansion provides the
fastest possible expansion rate (and therefore cooling rate) of the
quark-gluon plasma and thus enhances any nonequilibrium effects that
are based on the idea of a rapid quench. Since complete inhomogeneous
evolutions are at the edge of or beyond what is presently numerically
possible, this spherically symmetric expansion provides the other
extreme when compared with the slower, purely longitudinal expansion
studied earlier \cite {ref:cooperdcc}.

In terms of this coordinate system, Minkowski's line element
\begin{equation}
   ds^2 = dt^2 - dr^2 - r^2 (d\theta^2 + \sin^2 \theta d\varphi^2)
\end{equation}
is given by
\begin{equation}
   ds^2 = d\tau^2 - \tau^2(d\eta^2 + \sinh^2 \eta \; d\theta^2 +
      \sinh^2 \eta \sin^2 \theta \; d\varphi^2)
\>,
\end{equation}
from which we can read off the metric tensor
\begin{equation}
   g_{\mu \nu} = {\rm diag} \; (1, -\tau^2, -\tau^2 \sinh^2 \eta,
      -\tau^2 \sinh^2 \eta \sin^2 \theta)
\>.
\end{equation}
The metric in this coordinate system is of the Robertson-Walker form:
\begin{equation}
   g_{\mu \nu} = {\rm diag} \; (1, -{a^2(\tau)}, 
      -{a^2(\tau)} \sinh^2 \eta,
      -{a^2(\tau)} \sinh^2 \eta \sin^2 \theta)
\>,
\end{equation}
which corresponds to a Ricci flat cosmological model with uniform
expansion $a(\tau) = \tau$, and hyperbolic spatial sections, {\it
i.e.}, with curvature $\kappa = -1$.

It will be convenient to introduce the conformal time $u$ given by
\begin{equation}
   u = \int^{\tau} a^{-1}(\tau') \; d\tau'
\>,
\end{equation}
or equivalently
\begin{equation}
   \tau = \frac{{\rm e}^{u}}{m}
\>,
\end{equation}
where we choose $m = m_\pi$, the only mass scale in the system.  The
metric in the coordinate system with conformal time $(u, \eta, \theta,
\varphi)$ has the form
\begin{equation}
   g_{\mu \nu} = C(u)\; {\rm diag} \; (1, -1, -\sinh^2 \eta,
      -\sinh^2 \eta \sin^2 \theta)
\>,
\end{equation}
with 
\begin{eqnarray}
   C(u) & = & a^2(\tau) = \tau^2 = {\rm e}^{2u} / m^2,
   \nonumber \\
   \sqrt{-g} & = & C^2(u) \sinh^2 \eta \sin \theta
\>.
\end{eqnarray}

In a hydrodynamical model, all the expectation values depend only on
the proper time of the system. We therefore will assume that the mean
fields $\phi_i$ and $\chi$ are only functions of $u$, that is $\phi_i
= \phi_i(u)$ and $\chi = \chi(u)$. We then write the full quantum
field $\Phi_i$ in terms of its expansion about its mean value, as
follows:
\begin{equation}
   \Phi_i(u,\eta,\theta,\phi) = \phi_i(u) 
      + \hat\phi_i(u,\eta,\theta,\phi),
\end{equation}
where $\hat\phi_i$ are the quantum fluctuations, which include both
vacuum and thermal excitations.  The equations of motion are then
\begin{eqnarray}
   \left[ C^{-1}(u) 
      \left( \frac{{\partial}^2}{\partial u^2}
	+ 2 \frac{\partial}{\partial u}
	 \right) + \chi(u) \right]
      \phi_i(u) & = & H \delta_{i0},
   \nonumber \\
   \left[\Box + \chi(u) \right] \hat\phi_i(x) & = & 0
\>,
\end{eqnarray}
where $x$ is the four vector $x = (u, \eta, \theta, \varphi)$.
Then for $G_0(x,x';\chi)$ we find
\begin{displaymath}
   G_0(x,x';\chi) = i \; \langle {\rm T}_c \{
      \hat\phi_i(x), \hat\phi_i(x') \} \rangle
\>,
\end{displaymath}
where T$_c$ corresponds to a $u$-ordered product, following the
closed-time-path formalism of Schwinger \cite{ref:ctp}.

In order to solve the wave equation for the quantum fluctuations, we
follow Parker and Fulling \cite{ref:parker}, and write a mode
expansion for $\hat\phi_i$ as follows
\begin{equation}
   \hat{\phi}_i(u,\vec{x}) = C^{-1/2}(u) \; 
      \int_0^{\infty} d s
      \sum_{lm} \bigl[ 
      \hat{a}_{i,slm} \, g_s(u) {\cal{Y}}_{slm}(\vec{x}) +
      {{\hat{a}}^{\dag}}_{i,slm} 
      \, {g_s^*}(u) {\cal{Y}}^\ast_{slm}(\vec{x}) \bigr]
\>,
\label{eq:modes}
\end{equation}
with
\begin{displaymath}
   \vec{x} = (\eta, \theta, \phi),
\end{displaymath}
and
\begin{displaymath}
   \Delta^{(3)}{\cal Y}_{slm}(\vec{x})+ (s^2 + 1) 
      {\cal Y}_{slm}({\vec {x}})=0
\>,
\end{displaymath}
where $\Delta^{(3)}$ is the Laplacian of the three dimensional
hyperbolic spatial sections of curvature $\kappa = -1$.  Then the mode
functions $g_s(u)$ satisfy the following differential equation
\begin{equation}
   {\ddot{g}}_{s}(u) + \left[ s^2 +  C(u) \chi (u) \right] g_s(u)=0
\>,
\end{equation}
and for $G_0(x,x;\chi)$ we find \cite{ref:mypaper}
\begin{equation}
   \langle \hat\phi_i^2 \rangle = - i \; G_0(x,x;\chi) =
      C^{-1}(u) \int_{0}^{\infty} \, \frac{s^2 ds}{2{\pi^2}}
      (2n_s+1) |g_s(u)|^2
\>,
\end{equation}
once we have chosen a particular vector state with respect to which we shall
be taking expectation values.  We choose an initial state such that
the pair densities are zero, and the particle number density is
finitely integrable with respect to the corresponding integration
measure, namely
\begin{eqnarray*}
   \langle \hat{a}_{j,s'l'm'}^{\dagger} \hat{a}_{i,slm} \rangle & = &
      n_s \, \delta_{i j} \, \delta(s' - s) \, 
      \delta_{l l'} \, \delta_{m m'} \>, 
   \\ 
      \langle \hat{a}_{j,s'l'm'} \hat{a}_{i,slm}^{\dagger}
      \rangle & = & ( n_s + 1 ) \, \delta_{i j} \, \delta(s' - s) \,
      \delta_{l l'} \, \delta_{m m'} \>,
   \\
      \langle  {{\hat{a}}^{\dag}}_{i,slm} {{\hat{a}}^{\dag}}_{j,s'l'm'} 
      \rangle & = & p_s \, \delta_{ij} \, \delta(s' - s) \,
      \delta_{ll'} \, \delta_{mm'} = 0 \>,
   \\
      \langle  {{\hat{a}}}_{i,slm} {{\hat{a}}}_{j,s'l'm'} \rangle
      & = & p_s^{\ast} \, \delta_{ij} \, \delta(s' - s) \,
      \delta_{ll'} \, \delta_{mm'} = 0
\>.
\end{eqnarray*}
Here we choose the initial particle number density to be a thermal distribution
\begin{displaymath}
   n_s = [e^{\omega_s(u_0)/k_BT}-1]^{-1}
\>,
\end{displaymath}
where ${{\omega}^2_s(u_0)} = s^2 + C(u_0) \chi (u_0)$.  Notice that we
can choose the pair density to vanish  $(p_s = 0)$, 
since one has the freedom to
make a Bogoliubov transformation at $u_0$ so that this always remains
true. The boundary conditions on the mode functions $g_s(u)$ is that
they correspond to the positive frequency adiabatic mode functions
initially. We therefore choose
\begin{eqnarray}
   g_s(u_0) &=& 1 / \sqrt{2 \omega_s(u_0)},
   \nonumber \\
   \dot{g}_s(u_0) &=& - \left\{ 
      \frac{\dot{\omega}_s(u_0)}{2 \omega_s(u_0)}
      + i \omega_s(u_0) \right\} g_s(u_0)
\>.
\label{eq:initmodes}
\end{eqnarray}
The following rescalings will be useful:
\begin{mathletters}
\label{allrescale}
   \begin{equation}
      \phi_i(u) = C^{-1/2}(u) \rho_i(u),
      \label{eq:rho}
   \end{equation}
   \begin{equation}
      \chi(u) = C^{-1}(u) \tilde\chi(u),
      \label{eq:tildechi}
   \end{equation}
   \begin{equation}
      j_i = C^{-3/2}(u) \tilde j_i(u),
   \label{eq:tildej}
   \end{equation}
   \begin{equation}
      v = C^{-1/2}(u) \tilde v(u)
   \>.
   \label{eq:tildev}
   \end{equation}
\end{mathletters}
In terms of the scaled variables, the equations of motion can be written
\begin{eqnarray}
   \left[ \frac{d^2}{du^2} + s^2 + \tilde\chi(u) \right] g_s(u) 
      & = & 0,
   \\
   \left[ \frac{d^2}{du^2} -1 + \tilde\chi(u) \right] \rho_i(u) 
      & = & \tilde{j}_i(u)
\>,
\label{eq:motion}
\end{eqnarray}
with the gap equation being
\begin{equation}
   \frac{\tilde\chi(u)}{\lambda} = -{\tilde v}^2(u) + \rho^2_i(u) 
      + N \int_{0}^{\infty} \, \frac{s^2 ds}{2{\pi^2}}
      (2n_s+1) |g_s(u)|^2
\>.
\label{eq:gap}
\end{equation}

%
\subsection{Construction of the Stress-Energy Tensor}
\label{subsec:construct}
The energy-momentum tensor is defined by \cite{ref:BirDav}
\begin{eqnarray}
   T_{\mu\nu} & = & (1 - 2\xi)
      ( \nabla_{\mu} \Phi_i ) \, ( \nabla_{\nu} \Phi_i )
      + \left( 2 \xi - \frac{1}{2} \right) g_{\mu \nu} g^{\alpha \beta}
      ( \nabla_{\alpha} \Phi_i ) \, ( \nabla_{\beta} \Phi_i )
   \nonumber \\
      & - & 2 \xi \Phi_i (\nabla_{\mu} \nabla_{\nu} \Phi_i) 
      + \frac{1}{2} \xi g_{\mu \nu} \Phi_i \Box \Phi_i 
      + \frac{1}{2}(1 - 3\xi) \chi g_{\mu \nu} \Phi_i^2 
   \nonumber \\
      & - & g_{\mu \nu} \frac{\chi v^2}{2} - g_{\mu \nu} 
      \frac{\chi^2}{4 \lambda} 
      - g_{\mu \nu} \left(1 - \frac{3\xi}{2} \right) j_i \Phi_i
\>.
\end{eqnarray}
Equivalently, we can write this equation as \cite{ref:callan}
\begin{equation}
   T_{\mu\nu} =  
      ( \nabla_{\mu} \Phi_i ) \, ( \nabla_{\nu} \Phi_i )
      + \xi (g_{\mu \nu} \Box
      - \nabla_{\mu} \nabla_{\nu}) \Phi_i^2
      - g_{\mu \nu} \left( {\cal L}[\Phi_i,\chi] 
      + j_i \Phi_i \right)
\>,
\label{eq:tmunu}
\end{equation}
where ${\cal L}[\Phi_i,\chi]$ is given by Eq.~(\ref{eq:lag}).  We take
the expectation value of the energy-momentum tensor in the thermal
initial state chosen, and make use of the mode expansion given in
Eq.~(\ref{eq:modes}) to examine the components of $T_{\mu\nu}$.

After some algebra, we find for the $T_{uu}$ component
\begin{eqnarray}
   C(u)\langle {T}_{uu} \rangle & =  & 
      \frac{1}{2} \left[\dot\rho_i^2 +
      ( \tilde \chi + 1 - 12\xi ) \, \rho_i^2 \right] 
   \nonumber \\ &&
   + (6\xi - 1) \rho_i \dot\rho_i 
      - \tilde{j}_i \rho_i
      - \frac{\tilde \chi \tilde{v}^2}{2}
      - \frac{\tilde\chi^2}{4 \lambda} 
   \nonumber \\ &&  
   + \frac{N}{2} \int_0^{\infty} \frac{ s^2 d s }{ 2 \pi^2 }
        ( 2 n_s + 1 )
      \Biggl\{ 
      \left| \dot{g_s} \right|^2 +
      ( s^2 + \tilde \chi + 2 - 12\xi) \, | g_s |^2 
   \nonumber \\ &&
      + (6 \xi -1) (g_s \dot{g_s^\ast}
      + g_s^\ast \dot{g_s}) \Biggr\}
\>,
\label{eq:tuu}
\end{eqnarray}
and similarly for the $T_{\eta \eta}$ component
\begin{eqnarray}
   C(u) \langle {T}_{\eta\eta} \rangle & = & 
      \frac{1}{2} \left[ \dot\rho_i^2(1 - 4\xi) +
      ( 1 - \tilde \chi - 8\xi + 4\xi\tilde \chi ) \, \rho_i^2 \right]
   \nonumber \\ &&
      + (6\xi - 1) \rho_i \dot{\rho_i}
      + (1-2\xi) \tilde{j}_i \rho_i 
      + \frac{\tilde\chi \tilde{v}^2}{2}
      + \frac{{\tilde\chi}^2}{4 \lambda}
   \nonumber \\ &&
   + \frac{N}{2} \int_0^{\infty} \frac{ s^2 d s }{ 2 \pi^2 }
     ( 2 n_s + 1 )
      \Biggl\{
       (1 - 4\xi)\left| \dot{g_s} \right|^2
   \nonumber \\ &&
    + \left[-\frac{s^2}{3} - \tilde \chi + \frac{2}{3}
      + 4\xi(s^2 + \tilde\chi - 1) \right] 
      \, | g_s |^2
   \nonumber \\ &&
      + (6\xi - 1)(g_s \dot{ g_s^\ast }
      + g_s^\ast \dot{ g_s })\Biggr\}
\>.
\label{eq:tetaeta}
\end{eqnarray}

It can be easily shown that the energy-momentum tensor is diagonal and
that the spatial components are all equal, except for a geometrical
factor (see Appendix \ref{app:fluid}).  We then define
\cite{ref:cooperlandau}
\begin{equation}
   \langle \, T_{\mu\nu} \, \rangle \equiv C(u)
      {\rm diag} \, ( \, 
         \epsilon, \, p, \, 
         p \, \sinh^2 \eta , \, 
         p \, \sinh^2 \eta \sin^2 \theta \, ) 
\>.
\label{eq:diagtmunu}
\end{equation}
%
%
\section{Renormalization of $T_{\mu\nu}$}
\label{sec:renorm}
In order to analyze the divergences of $\langle T_{\mu \nu} \rangle$
we make use of the adiabatic expansion of the modes $g_s$, since we
know \cite{ref:bunch} that ${\langle T_{\mu \nu} \rangle}$ and
${\langle T_{\mu \nu} \rangle}_{\rm adiabatic}$ have the same
ultraviolet behaviour. Up to second adiabatic order we have
\cite{ref:bunch}
\begin{mathletters}
\label{alladiabatic}
   \begin{equation}
      g_s^A = \frac{1}{\sqrt{2 \Omega_s}} \; 
         {\rm exp} \left[ - i \int^u du' \Omega_s(u') \right] ,
   \label{eq:ga}
   \end{equation}
   \begin{equation}
      |g_s^A|^2 = \frac{1}{2\Omega_s} 
         = \frac{1}{2\omega_s} + \frac{\ddot \omega_s}{8\omega_s^4}
         - \frac{3 \dot\omega_s^2}{16 \omega_s^5} + \cdots,
   \label{eq:ga2}
   \end{equation}
   \begin{equation}
      |\dot g_s^A|^2 = \frac{1}{2\Omega_s}
         \left(\frac{\dot\Omega_s^2}{4\Omega_s^2} + \Omega_s^2 \right) 
         =  \frac{\omega_s}{2} 
         - \frac{\ddot \omega_s}{8\omega_s^2}
         + \frac{5 \dot\omega_s^2}{16 \omega_s^3} + \cdots,
   \label{eq:gadot2}
   \end{equation}
   \begin{equation}
      (g_s^A \dot g_s^{A \ast} + \dot g_s^A g_s^{A \ast}) = 
        - \frac{1}{2} \frac{\dot \Omega_s}{\Omega_s^2}
	=    - \frac{1}{2} \frac{\dot \omega_s}{\omega_s^2} + \cdots
   \label{eq:ggdot}
   \>,
   \end{equation}
\end{mathletters}
where $\omega_s^2 (u)= s^2 + \tilde \chi (u)$.

We regularize our integrals by introducing a non-covariant cutoff
$\Lambda$ in physical momentum, which corresponds to a comoving
momentum cutoff $s_m= C^{1/2}(u) \Lambda$. We shall work with a fixed
physical cutoff $\Lambda$, so that the comoving cutoff $s_m$ depends
on the conformal time $u$.

We have three physical parameters to renormalize: the mass of the pion
$m_\pi$, the self-interaction coupling constant $\lambda$, and the
(dimensionless) coupling constant to gravity $\xi$, necessary in order
for the field theory to be renormalizable, even in Ricci flat
spacetimes, such as the one considered in this paper
\cite{ref:toms}. The quadratic divergences in the gap equation will be
removed by mass renormalization, and the logarithmic divergences will
be subtracted by renormalization of the coupling constants $\lambda$
and $\xi$.  In addition, we shall see that there is an extra quartic
divergence in both the energy density and the pressure, coming from
the mode integrals in ${\langle T_{\mu \nu} \rangle}$ that carry an
extra factor of $k^2$, which can be removed by renormalizing the
cosmological constant.

First we examine the divergences in the gap equation
(\ref{eq:gap}). From the adiabatic expansion (\ref{alladiabatic}), we
know that the mode integral appearing in this equation has a quadratic
divergence. In order to remove this divergence, we perform mass
renormalization by subtracting the regularized gap equation for the
vacuum state:
\begin{equation}
   \frac{C(u) m^2}{\lambda} = -{\tilde v}^2(u) + C(u) f_\pi^2 
      + \frac{N}{2} \int_{0}^{s_m} \, \frac{s^2 ds}{2{\pi^2}}
      \frac{1}{{(s^2 + {\rm e}^{2u})}^{1/2}}
\>.
\label{eq:regvacgap}
\end{equation}
This yields for the gap equation
\begin{equation}
   \frac{\tilde \chi (u)}{\lambda} = \frac{C(u) m^2}{\lambda} 
      + \rho_i^2 (u) - C(u)  f_\pi^2 
      + N \int_{0}^{s_m} \, \frac{s^2 ds}{2{\pi^2}}
      \left[  (2n_s+1)  {|g_s(u)|}^2 -
      \frac{1}{2{(s^2 + {\rm e}^{2u})}^{1/2}} \right]
\>.
\label{eq:reggap}
\end{equation}
Note that the second integral is {\em independent} of time.  Once we
have removed the quadratic divergences in the gap equation, we are
only left with logarithmic divergences, which are subtracted by the
following coupling constant renormalizations
\begin{mathletters}
\label{eq:coupren}
   \begin{equation}
      \frac{1}{\lambda} = \frac{1}{\lambda_{\tiny R}} 
         - \frac{N}{4} \int_0^{s_m} \, \frac{ s^2 d s }{ 2 \pi^2 }
         \frac{1}{(s^2 + {\rm e}^{2u})^{3/2}} ,
   \label{eq:lambdaren}
   \end{equation}
   \begin{equation}
      \left( \xi - {1 \over 6} \right) {1 \over \lambda}
         = \left( \xi_{\tiny R} 
         - {1 \over 6} \right) {1 \over \lambda_{\tiny R}}
\>.
   \label{eq:xiren}
   \end{equation}
\end{mathletters}
Note that $\xi = 1/6 = \xi_{\tiny R}$ is a fixed point
of the renormalization flow equations \cite{ref:nonminimal}.

The renormalized form of the gap equation is then given by the
following
\begin{eqnarray}
   \frac{\tilde \chi(u)}{\lambda_{\tiny R}} 
      & = &
      - {\tilde v}^2_{\tiny R} 
      - \frac{NC(u)m^2}{16 \pi^2} + \rho_i^2
      + N \int_0^{s_m} \,  
      \frac{ s^2 d s }{ 2 \pi^2 } (2 n_s +1) {|g_s(u)|}^2 
   \nonumber \\ &&
      - \frac{NC(u)}{2} \int_0^{\Lambda} \,  
      \frac{ k^2 d k }{ 2 \pi^2 } \frac{1}{{(k^2 + m^2)}^{1/2}}
      - \frac{N[\tilde \chi(u) - C(u)m^2]}{4} \int_0^{\Lambda} \,  
      \frac{ k^2 d k }{ 2 \pi^2 } \frac{1}{{(k^2 + m^2)}^{3/2}}
\>.
\end{eqnarray}

We now proceed to write the regularized expressions for the energy and
isotropic pressure and analyze the divergences appearing in the mode
integrals.  We shall take $\xi_{\tiny R} = 1/6$ as suggested by
\cite{ref:nonminimal}. The physics behind this choice is clear.  If
one considers an arbitrary composite scalar field (such as the meson
field considered here) and an effective field theory at a scale
$\Lambda$, and one carries out a renormalization group analysis in
the leading large-$N$ approximation (or a fully improved one-loop
renormalization group approximation), $\xi = 1/6 = \xi_{\tiny R}$ is
found to be an attractive renormalization group fixed point in the
infrared limit \cite{ref:nonminimal}.  This means that even if there are
corrections to $\xi = 1/6$ at large scales (of order $\Lambda$), the
observed low energy value of the coupling $\xi$ tends to a physical
value of $\xi_{\tiny R} = 1/6$, as one evolves into the infrared.  The
choice $\xi = 1/6 = \xi_{\tiny R}$ also implies that the divergences
of both the energy and the pressure can be obtained at adiabatic order
zero \cite{ref:bunch}.

Now we examine the divergences present in the mode integrals of 
the energy density and pressure. Recall that
\begin{eqnarray}  
   {\cal E}(u) & \equiv &   C^2(u) \epsilon(u)
      \equiv  \frac{1}{2} \left[ \dot\rho_i^2 
      + \rho_i^2 (\tilde\chi - 1) \right] 
      - \tilde{j}_i \rho_i
      - \frac{\tilde\chi \tilde{v}^2}{2}
      - \frac{\tilde\chi^2}{4 \lambda}
   \nonumber \\ &&
      + \frac{N}{2} \int_0^{s_m} \frac{ s^2 d s }{ 2 \pi^2 }
      ( 2 n_s + 1 ) \{ 
      \left| \dot{g_s} \right|^2 
      + ( s^2 + \tilde\chi ) \, | g_s |^2 \}
\label{eq:regepsu}
\>,
\end{eqnarray}
and
\begin{eqnarray}
   {\cal P}(u) & \equiv & C^2(u) p(u)
      \equiv  \frac{1}{6} \left[ \dot\rho_i^2 
      - \rho_i^2 (\tilde\chi + 1) \right] 
      + \frac{2}{3} \tilde{j}_i \rho_i 
      + \frac{\tilde\chi \tilde{v}^2}{2}
      + \frac{\tilde\chi^2}{4 \lambda}  
  \nonumber \\ &&
      + \frac{N}{6} \int_0^{s_m} \frac{ s^2 d s }{ 2 \pi^2 }
      ( 2 n_s + 1 ) \{ 
      \left| \dot{g_s} \right|^2 
      + ( s^2 - \tilde\chi ) \, | g_s |^2 \}
  \label{eq:regpu}
\>.
\end{eqnarray}
For the energy the contribution coming from the mode integrals is
\begin{equation}
   \frac{N}{2} \int_0^{s_m} \frac{ s^2 d s }{ 2 \pi^2 }
      ( 2 n_s + 1 ) \left\{ 
      \left| \dot{g}_s \right|^2 + 
      ( s^2 + \tilde\chi ) \, | g_s |^2 \right\} 
\>,
\end{equation}
and using the adiabatic expansion given in Eq.~(\ref{alladiabatic}),
the divergent part of this integral is given by
\begin{eqnarray}
   C^2(u) \epsilon_0 & \equiv & 
      \frac{N}{2} \int_0^{s_m} \frac{s^2 d s}{2 \pi^2} \, \omega_s =
      \frac{N}{2} \int_0^{s_m} \frac{s^2 d s}{2 \pi^2} \,
      \sqrt{s^2 + \tilde\chi}
   \nonumber \\ &=& 
      \frac{N s_m^4}{16\pi^2} + \frac{N s_m^2 \tilde\chi}{16 \pi^2} 
      - \frac{N \tilde\chi^2}{64 \pi^2} \ln 
      \left( \frac{4s_m^2}{\tilde\chi} \right) 
      + \frac{N \tilde\chi^2}{128 \pi^2} + {\cal O}
      \left(\frac{\tilde\chi^3}{s_m^2}\right) 
\>.
\label{eq:edivergent}
\end{eqnarray}
For the pressure the contribution coming from the mode integrals is
\begin{equation}
   \frac{N}{6} \int_0^{s_m} \frac{ s^2 d s }{ 2 \pi^2 }
      ( 2 n_s + 1 ) \left\{
      \left| \dot{g}_s \right|^2 + 
      (s^2 - \tilde\chi) \, | g_s |^2 \right\}
\>.
\end{equation}
Again using the adiabatic mode expansion, the divergent part of this
integral is given by
\begin{eqnarray}
   C^2(u) p_0 & \equiv &
      \frac{N}{6} \int_0^{s_m} \frac{ s^4 d s }{ 2 \pi^2 } 
      \frac{1}{\sqrt{s^2 + \tilde\chi}}
   \nonumber\\
   &=& \frac{N s_m^4}{48 \pi^2} - \frac{N s_m^2 \tilde\chi}{48 \pi^2}
      + \frac{N \tilde\chi^2}{64 \pi^2} 
      \ln \left( \frac{4s_m^2}{\tilde\chi} \right)
      - \frac{7N \tilde\chi^2}{384 \pi^2}
      + {\cal O} \left(\frac{\tilde\chi^3}{s_m^2}\right) 
\>.
\label{eq:pdivergent}
\end{eqnarray}

Notice that $\epsilon_0 \ne - p_0$, as is required by general
covariance (see Appendix {\ref{app:christensen}}).  We must then
enforce covariance {\em by hand}.  We mention here that the
introduction of a momentum cutoff does not spoil covariance at the
level of the energy density (see Appendix {\ref{app:christensen}}). On
the other hand, we must carefully handle the subtractions that will
lead to the finite isotropic pressure \cite{ref:phi4}.  The quartic
subtraction in the energy density and pressure is a renormalization of
the cosmological constant; in other words, a subtraction of the
cosmological vacuum energy ${N \Lambda^4} / {16 \pi^2}$.

If we now define
\begin{mathletters}
\label{eq:e-prenormalized}
   \begin{equation}
      \epsilon_{\tiny R} = \epsilon
         - \frac{N s_m^4}{16\pi^2 C^2(u)},
      \label{eq:eren}
   \end{equation}
   \begin{equation}
      p_{\tiny R} = p - p_0 - \epsilon_0 
      + \frac{N s_m^4}{16\pi^2 C^2(u)}
      \label{eq:pren}
\>,
   \end{equation}
\end{mathletters}
we can easily show that the energy density and pressure are finite, by
making use of Eqs.~(\ref{eq:regvacgap}) and (\ref{eq:coupren}).

In order to numerically evaluate the physical energy density and
pressure, we must also make sure that the vacuum energy that we are
measuring with respect to is zero ({\em i.e.} when we are at the
minimum of the potential). This vacuum energy is calculated in the
``out'' regime of the collision, when $\langle \sigma \rangle =
f_\pi$, $\langle \vec\pi_i \rangle = 0$, $\chi = m_\pi^2$, and the
mode functions are the zeroth order adiabatic ones with $\omega_s^2 =
s^2 + C(u)m^2$.  The renormalized vacuum energy is given by
\begin{equation}
   \epsilon_{\tiny {\rm vac}} =  \frac{m^2 f_\pi^2}{2}
      - \frac{m^2 v^2}{2} - \frac{m^4}{4\lambda}
      - \frac{N \Lambda^4}{16 \pi^2}
      + \frac{N}{2 C^2} \int_0^{s_m} \, \frac{s^2 ds}{2\pi^2}
      \sqrt{s^2 + m^2 C}
\>.
\end{equation}
The vacuum value of the pressure is $p_{\tiny {\rm vac}} = -
\epsilon_{\tiny {\rm vac}}$. These values must be subtracted from
Eqs.~(\ref{eq:e-prenormalized}) so that when we have reached the
``out'' regime, the energy density and pressure go to zero.

We calculate now the trace of the energy-momentum tensor, and show
that these definitions of $\epsilon_{\tiny R}$ and
$p_{\tiny R}$ give a finite trace. We have from its definition
\begin{equation}
   - \langle T_{\mu}^{\mu} \rangle_{\tiny R }
      = 3 p_{\tiny R} - \epsilon_{\tiny R}
\>.
\label{eq:e-mtrace}
\end{equation}
If we make use of Eqs.~(\ref{eq:e-prenormalized}), we can write
\begin{equation}
   - C^2(u) \, \langle T_{\mu}^{\mu} \rangle_{\tiny R} = 
      \tilde{v}_{\tiny R}^2 \tilde{\chi}(u)
      + 3 \tilde{j}_i \rho_i(u)
      + \frac{N \tilde{\chi}^2(u)}{32 \pi^2}
\>,
\label{eq:trace}
\end{equation}
where we have defined
\begin{equation}
   \tilde{v}^2 =
      - \frac{m^2 C(u)}{\lambda_{\tiny R}} + f_\pi^2 C(u)
      -\frac{N m^2 C(u)}{16 \pi^2}
      + \frac{N s_m^2}{8 \pi^2} \equiv
      \tilde{v}_R^2 + \frac{N s_m^2}{8 \pi^2} 
\>.
\label{eq:v2r}
\end{equation}

The first two terms in Eq.~(\ref{eq:trace}) are the renormalized
classical trace of the energy-momentum tensor, and the second term is
the one-loop quantum trace anomaly.

%
\section{Numerical Results}
\label{sec:results}
\subsection{Energy-Momentum Tensor Conservation}
\label{sec:engcons}
The renormalized energy-momentum tensor obeys the conservation law
\begin{equation}
   \nabla_{\mu} {\langle T^{\mu\nu} \rangle}_{\tiny R}  = 0
\>.
\label{eq:dmutmunu}
\end{equation}

Using the Christoffel symbols tabulated in Appendix~\ref{app:chris},
we find that the $\nu = u$ component of the conservation equation
takes the form
\begin{equation}
   \dot\epsilon_{\tiny R}(u) + 3 [p_{\tiny R}(u) 
      + \epsilon_{\tiny R}(u)] = 0  
\>.
\label{eq:uecons}
\end{equation}

In terms of the variables
\begin{mathletters}
\label{eq:e-pcal}
   \begin{equation}
      {\cal E}_{\tiny R} = {\cal E}
         - \frac{N s_m^4}{16\pi^2},
      \label{eq:ecalren}
   \end{equation}
   \begin{equation}
      {\cal P}_{\tiny R} = {\cal P} - C^2(u) ( p_0 + \epsilon_0) 
      + \frac{N s_m^4}{16\pi^2}
      \label{eq:pcalren}
\>,
   \end{equation}
\end{mathletters}
we can rewrite the conservation equation as
\begin{equation}
   \dot{\cal E}_{\tiny R}(u) + 3 {\cal P}_{\tiny R}(u) 
      - {\cal E}_{\tiny R}(u) = 0  
\>.
\end{equation}

Recall that
\begin{eqnarray}  
   {\cal E}_{\tiny R} & = & \frac{1}{2} \left[ \dot\rho_i^2 
      + \rho_i^2 (\tilde\chi - 1) \right] 
      - \tilde{j}_i \rho_i
      - \frac{\tilde\chi \tilde{v}^2}{2}
      - \frac{\tilde\chi^2}{4 \lambda}
   \nonumber \\ &&
      + \frac{N}{2} \int_0^{s_m} \frac{ s^2 d s }{ 2 \pi^2 }
        ( 2 n_s + 1 ) \{ 
      \left| \dot{g_s} \right|^2 
      + ( s^2 + \tilde\chi ) \, | g_s |^2 \}
      - \frac{N s_m^4}{16\pi^2}
\>.
\end{eqnarray}
Therefore
\begin{eqnarray}
   \dot{\cal E}_{\tiny R}
      &=& \dot\rho_i \ddot \rho_i 
      + (\tilde\chi - 1)\rho_i \dot\rho_i
      + \frac{1}{2} \dot{\tilde\chi} \rho_i^2 
      - \frac{\dot{\tilde\chi} \tilde{v}^2}{2}
      - \tilde\chi \tilde{v} \dot{\tilde{v}}
      - \frac{\tilde\chi \dot{\tilde\chi}}{2\lambda}
      - (\dot{\tilde{j}}_i \rho_i + \tilde{j} \dot\rho_i)
   \nonumber \\
   &+& \frac{N}{2} \int_0^{s_m} \, \frac{s^2 ds}{2\pi^2}
      (2 n_s + 1) \left\{ \dot g_s \ddot g_s^\ast + \ddot g_s \dot g_s^\ast
      + (s^2 + \tilde\chi) (g_s \dot g_s^\ast + \dot g_s g_s^\ast)
      + \dot{\tilde\chi} |g_s|^2 \right\}
   \nonumber \\
   &+& \frac{N}{2} \frac{s_m^3}{2\pi^2}
      (2 n_{s_m} + 1) \left\{ |\dot g_{s_m}|^2 
         + (s_m^2 + \tilde\chi)|g_{s_m}|^2 \right\}
         - \frac{N s_m^4}{4\pi^2}
\>,
\label{eq:erdot}
\end{eqnarray}
where we have taken into account the contribution coming from the
upper limit of the mode integral, since $s_m$ depends on $u$.  Making
use of the identities ${\dot{\tilde{v}}}_i = {\tilde{v}}_i$,
${\dot{\tilde{j}}}_i = 3{\tilde{j}}_i$, the equations of motion, and
using the adiabatic approximation for $g_{s_m}$ (we can always choose
$s_m$ to be a high enough comoving momentum so that the adiabatic
approximation is valid in this limit), we then obtain
\begin{equation}
   \dot{\cal E}_{\tiny R} = 
      - \tilde\chi \tilde{v}^2 
      - 3 \tilde{j}_i \rho_i
      + \frac{N}{4\pi^2} s_m^3 (2 n_{s_m} + 1) \omega_{s_m}
      - \frac{N s_m^4}{4\pi^2} 
\>.
\end{equation}
The occupation number $n_{s_m}$ goes to zero for large $s_m$, so we can
neglect it. Expanding out $\omega_{s_m}$ yields
\begin{equation}
   \omega_{s_m} = \sqrt{s_m^2 + \tilde\chi} = s_m
      \left(1 + \frac{\tilde\chi}{2 s_m^2}
      - \frac{\tilde\chi^2}{8 s_m^4} + \cdots \right)
\>.
\end{equation}
So that we have
\begin{equation}
   \dot {\cal E}_{\tiny R} = 
      - \tilde\chi \tilde{v}^2 + \frac{N \tilde\chi s_m^2}{8 \pi^2}     
	  - 3 \tilde{j}_i \rho_i     
	  - \frac{N \tilde\chi^2}{32 \pi^2}
\>,
\end{equation}
or equivalently
\begin{equation}
   \dot {\cal E}_{\tiny R} = 
      - \tilde\chi \tilde{v}_{\tiny R}^2
      - 3 \tilde{j}_i \rho_i 
      - \frac{N \tilde{\chi}^2}{32 \pi^2}
\>.
\end{equation}
If we compare this expression with Eq.~({\ref{eq:trace}}), we can
easily see that the conservation equation is satisfied.

%
\subsection{Effective Equation of State}
\label{sec:eos}

Now that we have finite equations for the energy density and pressure,
we can evolve them in proper time and numerically investigate their
behavior.  Since we have a nonequilibrium situation, it does not
really make sense to calculate an actual equation of state, {\em i.e.}
$p = p(\epsilon)$. However, since $\epsilon$ and $p$ are both
functions of $u$, in regions where both are monotonically increasing
or decreasing, one can determine an ``effective'' equation of state $p
= p(\epsilon)$. In this calculation, we do not have two-body
scattering, so that there are many oscillations in the pressure. We
expect at the next order in the $1/N$ expansion, when these effects
are included that one will be able to extract an effective equation of
state.

\subsection{Numerical Simulations}
We choose the initial state at a temperature above the phase
transition in thermal equilibrium, with all particle masses
positive. The equations are solved self-consistently at the starting
time to obtain the values of the mean fields. We fixed the value of
$\chi$ at the initial time as the solution of the gap equation in the
initial thermal state. We also required that the initial expectation
values of the $\sigma$ and $\vec\pi$ fields satisfy ${\vec \pi}^2(u_0) +
\sigma^2(u_0) = \sigma^2_{\tiny T}$, where $\sigma_{\tiny T}$ is the
thermal equilibrium value of $\Phi$ at the initial temperature
$T$. The critical temperature is $T_c \approx$ 160 MeV, so we choose
$T =$ 200 MeV, which gives $\sigma_{\tiny T} = 0.3$ fm $^{-1}$. The
mode functions are chosen as their adiabatic values [see
Eq.~(\ref{eq:initmodes})]. We choose the coupling constant
$\lambda_{\tiny R} = 7.3$ \cite{ref:mypaper}, and take the conformal
value of the coupling to gravity $\xi = \xi_{\tiny R} = 1/6$
\cite{ref:nonminimal}.

\begin{figure}
\epsfxsize = 4.0in
\centerline{\epsfbox{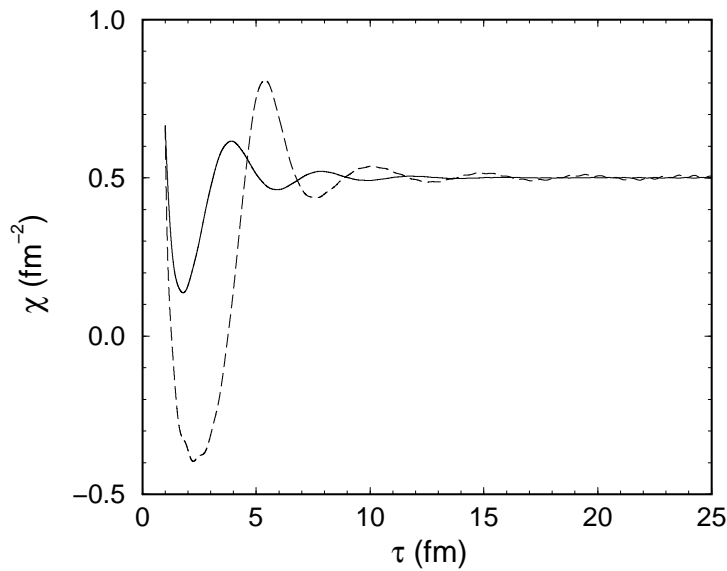}}
\caption{Proper time evolution of the $\chi$ field 
for the following initial conditions. 
Solid line: $\sigma(\tau_0) = \sigma_{\tiny T}$,
$\pi_i(\tau_0) = 0$, and $\dot\sigma(\tau_0) = 0$.
Dashed line: $\sigma(\tau_0) = \sigma_{\tiny T}$,
$\pi_i(\tau_0) = 0$, and $\dot\sigma(\tau_0) = -1$ }
\label{fig:chi}
\end{figure}

\begin{figure}
\epsfxsize = 4.0in
\centerline{\epsfbox{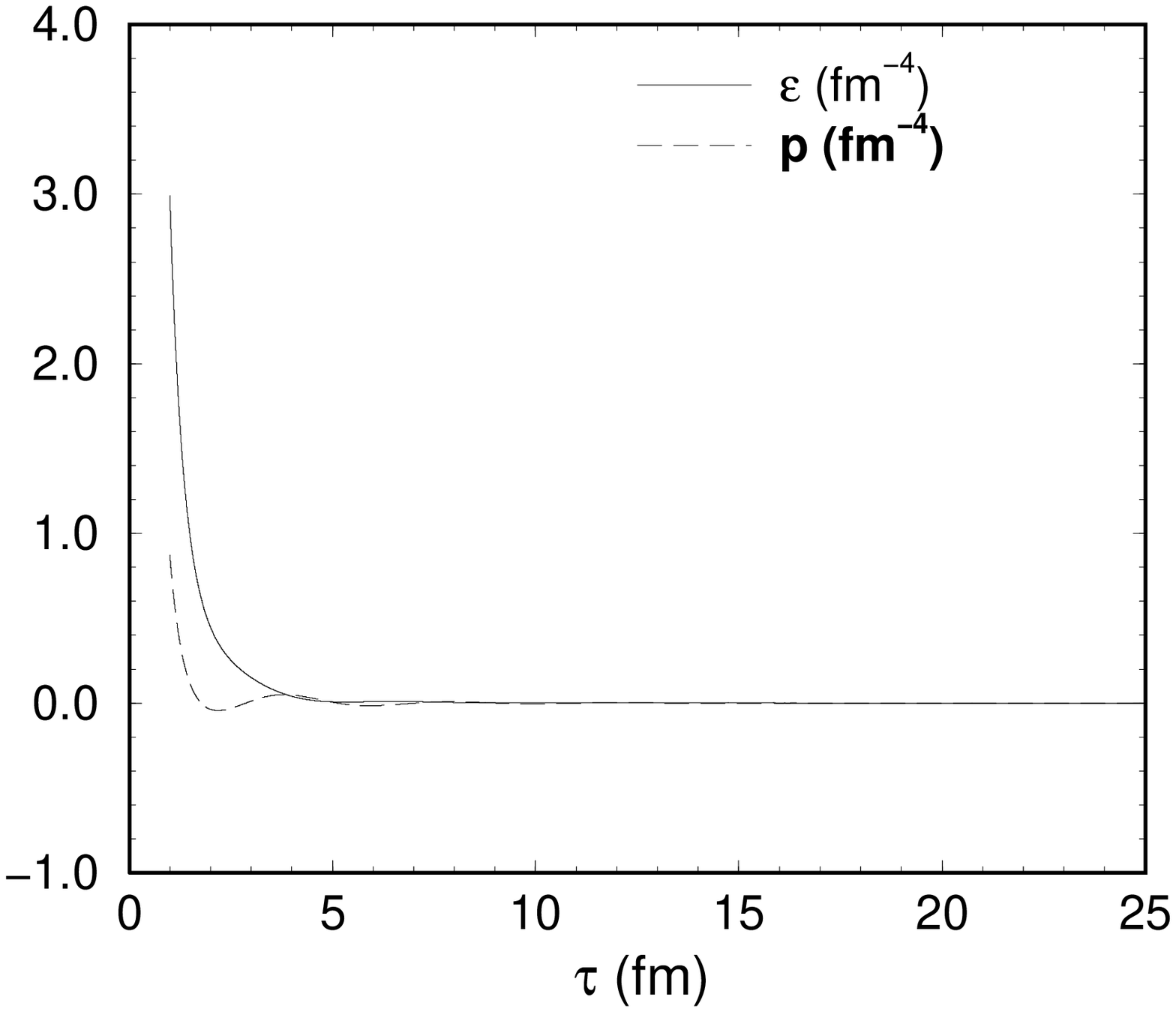}}
\caption{Proper time evolution of the energy density and pressure for
the initial conditions $\sigma(\tau_0) = \sigma_{\tiny T}$,
$\pi_i(\tau_0) = 0$, and $\dot\sigma(\tau_0) = 0$.}
\label{fig:eandp0}
\end{figure}

\begin{figure}
\epsfxsize = 4.0in
\centerline{\epsfbox{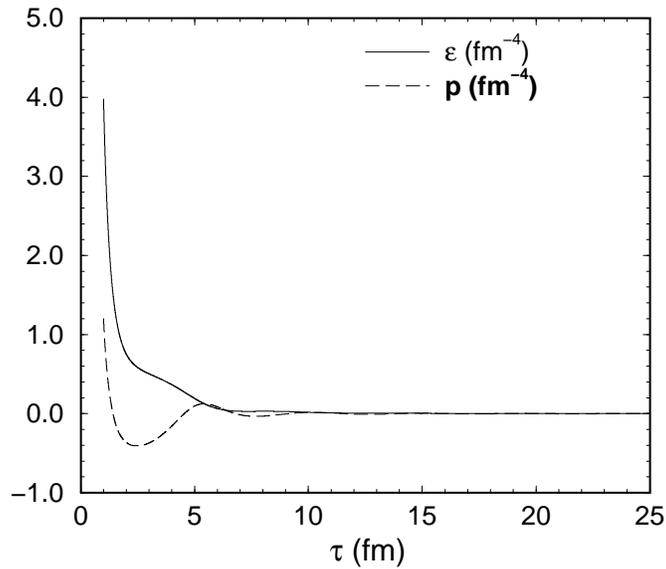}}
\caption{Proper time evolution of the energy density and pressure for
the initial conditions $\sigma(\tau_0) = \sigma_{\tiny T}$,
$\pi_i(\tau_0) = 0$, and $\dot\sigma(\tau_0) = -1$.}
\label{fig:eandpneg1}
\end{figure}

In Fig.~\ref{fig:chi}, we show the time evolution of the $\chi$ field,
which is the effective mass squared for the $\phi$ field. When this
field becomes negative, long-wavelength unstable modes begin to grow
exponentially. Therefore this field serves as a measure of instability
in the system. For initial conditions with a negative derivative, we
find that the $\chi$ field is negative for about 3 fm/$c$ of proper
time.  Figure \ref{fig:eandp0} shows the numerical calculation of the
energy density and pressure, in units of fm$^{-4}$, for initial
conditions where no instabilities arose. Note that the system quickly
reaches the ``out'' regime and relaxes to its vacuum values.  Figure
\ref{fig:eandpneg1} shows the same plot for initial conditions with an
instability. For the unstable initial conditions more energy density
was produced initially, but both the energy density and the pressure
drop off very quickly, as is expected (since the system is cooling
very rapidly). It is interesting to note that the pressure becomes
negative when $\chi$ is negative. If we define the speed of sound
$c_0^2(\tau)$ by $p = c_0^2(\tau) \epsilon$, we notice that $c_0^2 \leq 1$.

\begin{figure}
\epsfxsize = 4.0in
\centerline{\epsfbox{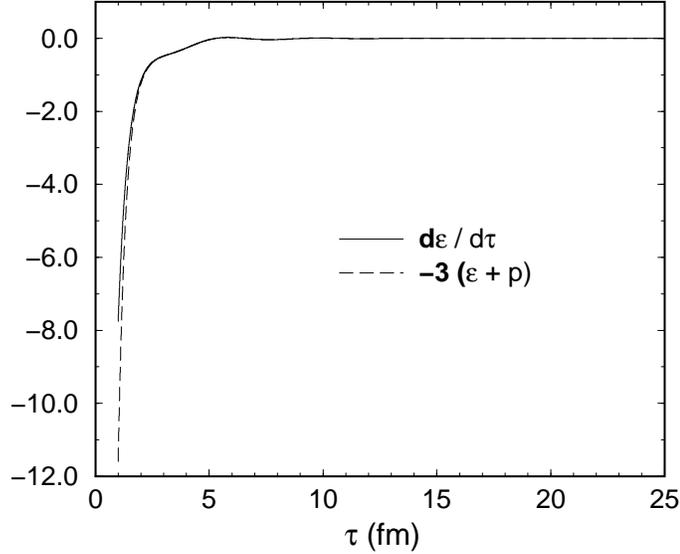}}
\caption{Energy-momentum tensor conservation, for the initial conditions
$\sigma(\tau_0) = \sigma_{\tiny T}$, $\pi_i(\tau_0) = 0$, and
$\dot\sigma(\tau_0) = 0$.}
\label{fig:cons0}
\end{figure}

\begin{figure}
\epsfxsize = 4.0in
\centerline{\epsfbox{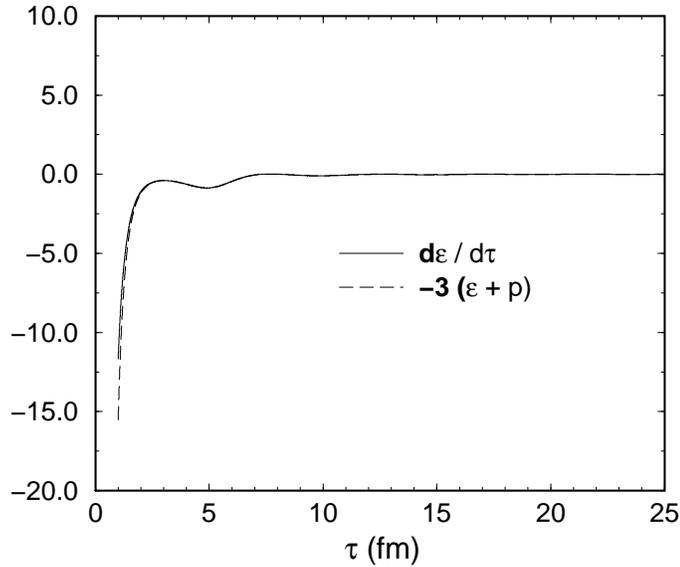}}
\caption{Energy-momentum tensor conservation, for the initial conditions
$\sigma(\tau_0) = \sigma_{\tiny T}$, $\pi_i(\tau_0) = 0$, and
$\dot\sigma(\tau_0) = -1$.}
\label{fig:consneg1}
\end{figure}

Figure \ref{fig:cons0} shows the conservation of the energy-momentum
tensor. Notice that it
is not conserved for short proper times. Because of the
presence of the Landau pole it is not possible to take the physical
cutoff $\Lambda$ very large (we use 800 MeV). Due to this rather small
cutoff, the occupation number $n_{s_m}$ is not actually zero, which it
was assumed to be for the derivation of the energy-momentum
 conservation [see the
discussion after Eq.~(\ref{eq:erdot})]. As soon as the cutoff $s_m$
becomes large enough, then the occupation number goes to zero and
the energy-momentum tensor is conserved. Figure \ref{fig:consneg1} shows 
the energy-momentum tensor
conservation for initial conditions with an instability. We see a
qualitatively similar behavior, regardless of initial conditions.

%
\section{Conclusions}
\label{sec:conc}
We have shown how to regularize and renormalize the energy-momentum
tensor for a spherically symmetric model with a time-dependent
comoving momentum cutoff which breaks covariance. We computed the
finite energy density and pressure, and numerically examined the
equation of state and proved energy conservation for this system. We
found that the energy density and pressure both decrease quickly,
which is expected for a rapidly cooling system. An interesting feature
of the expansion is that when the effective mass squared is negative,
the pressure is also negative. A negative pressure implies cavitation,
which could mean the formation of domains, {\em i.e.} DCCs.

%
\section*{Acknowledgements}
The authors would like to thank Yuval Kluger, Emil Mottola, and John
Dawson for useful discussions; and give special thanks to Fred Cooper
and Salman Habib for their contributions. UNH gratefully acknowledges
support by the U.S. Department of Energy (DE-FG02-88ER40410). One of
us (MAL) would like to thank Los Alamos National Laboratory for its
hospitality.

%
\appendix
\section{Christoffel symbols}
\label{app:chris}

The only non-vanishing Christoffel symbols for the metric
\begin{equation}
   g_{\mu \nu} = C(u)\; {\rm diag} \; (1, -1, -\sinh^2 \eta,
      -\sinh^2 \eta \sin^2 \theta)
\>,
\end{equation}
with 
\begin{eqnarray}
   C(u) & = & a^2(\tau) = \tau^2 = {\rm e}^{2u} / m^2
\>,
\end{eqnarray}
are
\begin{displaymath}
\begin{array}{rclrcl}
\Gamma^{u}_{\eta\eta} & = & 1 \>, &
\Gamma^{u}_{\theta\theta} & = & \sinh^2 \eta \>, \\
\Gamma^{u}_{\varphi\varphi} & = & \sinh^2 \eta \sin^2 \theta \>, & 
\Gamma^{\eta}_{\theta\theta} & = & -\sinh\eta \cosh\eta \>, \\
\Gamma^{\eta}_{\varphi\varphi} & = & -\sinh\eta \cosh\eta \sin^2 \theta \>, &
\Gamma^{\theta}_{\varphi\varphi} & = & -\sin\theta \cos\theta \>,
\end{array}
\end{displaymath}
\begin{displaymath}
\begin{array}{rcccccccl}
\Gamma^{u}_{uu} & = & \Gamma^{\eta}_{\eta u} 
  & = & \Gamma^{\theta}_{\theta u} 
  & = & \Gamma^{\varphi}_{\varphi u} & = & 1 \>,
\end{array}
\end{displaymath}
\begin{displaymath}
\begin{array}{rcccl}
\Gamma^{\theta}_{\theta\eta} & = & \Gamma^{\varphi}_{\varphi\eta} 
   & = & \coth\eta \>, 
\end{array}
\end{displaymath}
\begin{displaymath}
\begin{array}{rcl}
\Gamma^{\varphi}_{\varphi\theta} & = & \cot\theta \>.
\end{array}
\end{displaymath}

The Christoffel symbols are used to derive the conservation law for
the energy-momentum tensor. 

\section{Energy-momentum Tensor for a Perfect Fluid}
\label{app:fluid}
For a perfect fluid we can write the energy-momentum tensor as follows
\begin{equation}
   T_{\mu \nu} = - p \; g_{\mu \nu} + (\epsilon + p) \; u_\mu u_\nu
\>,
\end {equation}
where $u_\mu$ are the components of the velocity vector of the fluid,
such that it is normalized, {\em i.e.} $g^{\mu \nu} u_\mu u_\nu = 1$,
which corresponds to a timelike vector field. In the reference frame
where the fluid is at rest we can write
\begin{equation}
   u_\mu = (u_u, u_\eta, u_\theta, u_\varphi) = (u_u, \vec{0})
\>,
\end{equation}
and the normalization condition simply implies that $g_{uu} = u_u u_u
= C(u)$.  It is then straightforward to obtain the components of the
energy-momentum tensor in this coordinate system
\begin{eqnarray}
   T_{uu} & = & g_{uu} \; \epsilon = C(u)\; \epsilon,
   \nonumber \\
   T_{\eta \eta} & = & - g_{\eta \eta} \; p = C(u) \;  p,
   \nonumber \\
   T_{\theta \theta} & = & - g_{\theta \theta}\;  p 
      = C(u) \; p \sinh^2 \eta,
   \nonumber\\
   T_{\varphi \varphi} & = & - g_{\varphi \varphi} \;  p = 
      C(u) \; p \sinh^2 \eta \sin^2 \theta
\>.
\end{eqnarray}
Then we can write
\begin{equation}
   \langle \, T_{\mu\nu} \, \rangle \equiv C(u)
      {\rm diag} \, ( \, 
      \epsilon, \, p, \, 
      p \, \sinh^2 \eta , \, 
      p \, \sinh^2 \eta \sin^2 \theta \, ) 
\>.
\end{equation}

%
\section{Regularization by Covariant Geodesic Point-Splitting}
\label{app:christensen}
The regularization scheme (introducing a physical momentum cutoff
$\Lambda$) used to render the mode integrals finite in the energy
density and isotropic pressure is not a covariant method.  Since we
must obtain covariant results (we want to obtain $\langle T_{\mu
\nu}\rangle_{\rm renormalized}$), we have to proceed with care, in
order to perform the physical subtractions that will yield the
physical finite energy density and pressure.  We shall make use of the
results obtained by Christensen \cite{ref:christensen}, in order to
analyze the non-covariant structure on the divergences in $\epsilon_0$
and $p_0$, since we know that covariant point-splitting in a spatial
direction is equivalent to regularization by introducing a momentum
cutoff.

Christensen \cite{ref:christensen} obtained:
\begin{eqnarray}
   \langle T_{\mu \nu} \rangle_{quartic}  & = &
      {\lim}_{x' \rightarrow x} \frac{1}{2 \pi^2} \frac{1}
      {{(\sigma^\lambda \sigma_\lambda)}^2}
      \left( g_{\mu \nu} - \frac{4 \sigma_\mu \sigma_\nu}
      {{(\sigma^\lambda \sigma_\lambda)}} \right),
   \nonumber \\
   \langle T_{\mu \nu} \rangle_{quadratic} & = & -
      {\lim}_{x' \rightarrow x} \frac{1}{4 \pi^2} \frac{1}
      {{(\sigma^\lambda \sigma_\lambda)}}
      \frac{\chi}{2}
      \left( g_{\mu \nu} - \frac{2 \sigma_\mu \sigma_\nu}
      {{(\sigma^\lambda \sigma_\lambda)}}
      \right),
	\nonumber   \\
   \langle T_{\mu \nu} \rangle_{logarithmic} & = & -
      {\lim}_{x' \rightarrow x} \frac{1}{4 \pi^2}
      \frac{\chi^4}{8} g_{\mu \nu}
      \left( \gamma + \frac{1}{2} \ln \mid\frac{\chi \sigma}{2} \mid
      \right)
\>.
\end{eqnarray}

If we choose $\sigma_\mu$ to be a spacelike vector such that
\begin{equation}
   \sigma_i \sigma_j = \frac{1}{3} g_{ij} (\sigma^\lambda \sigma_\lambda)
      \; \; \; \; {\rm and} \; \; \; \;  \sigma_u =0 \; \; \; \; 
      {\rm with} \; \; \; \; i, j = \eta, \theta, \varphi
\>,
\end{equation}
we can write
\begin{eqnarray}
   \langle T_{uu} \rangle_{quartic}  & = &
      {\lim}_{x' \rightarrow x} \frac{1}{2 \pi^2} \frac{1}
      {{(\sigma^\lambda \sigma_\lambda)}^2} \;  C(u),
   \nonumber \\
   \langle T_{\eta \eta} \rangle_{quartic}  & = &
      {\lim}_{x' \rightarrow x} \frac{1}{2 \pi^2} \frac{1}
      {{(\sigma^\lambda \sigma_\lambda)}^2} \; \frac{C(u)}{3},
   \nonumber \\
   \langle T_{uu} \rangle_{quadratic} & = & -
      {\lim}_{x' \rightarrow x} \frac{1}{4 \pi^2} \frac{1}
      {(\sigma^\lambda \sigma_\lambda)} \; 
      \frac{\chi}{2} \; C(u),
   \nonumber \\
   \langle T_{\eta \eta} \rangle_{quadratic} & = & 
      {\lim}_{x' \rightarrow x} \frac{1}{4 \pi^2} \frac{1}
      {{(\sigma^\lambda \sigma_\lambda)}} \; 
      \frac{\chi}{2} \;  \frac{C(u)}{3},
	\nonumber \\
      \langle T_{uu} \rangle_{logarithmic} & = & -
      {\lim}_{x' \rightarrow x} \frac{1}{4 \pi^2} \;
      \frac{\chi^4}{8} \;  C(u)
      \left( \gamma + \frac{1}{2} \ln \mid\frac{\chi \sigma}{2} \mid
      \right),
   \nonumber \\
   \langle T_{\eta \eta} \rangle_{logarithmic} & = & 
      {\lim}_{x' \rightarrow x} \frac{1}{4 \pi^2}\; 
      \frac{\chi^4}{8} \; C(u)
      \left( \gamma + \frac{1}{2} \ln \mid\frac{\chi \sigma}{2} \mid
      \right)
\>.
\label{eq:e-mcovariant}
\end{eqnarray}

>From Appendix \ref{app:fluid} we know that $T_{uu} = C(u) \epsilon$
and that $T_{\eta \eta} = C(u) p$ and from the previous equations
(\ref{eq:e-mcovariant}) we can make the following identifications:
\begin{eqnarray}
   \epsilon_{quartic}  & = &
      {\lim}_{x' \rightarrow x} \frac{1}{2 \pi^2} \frac{1}
      {{(\sigma^\lambda \sigma_\lambda)}^2},
  \nonumber \\
   p_{quartic}  & = &
      {\lim}_{x' \rightarrow x} \frac{1}{2 \pi^2} \frac{1}
      {{(\sigma^\lambda \sigma_\lambda)}^2} \; \frac{1}{3},
   \nonumber \\
   \epsilon_{quadratic} &  = & -
      {\lim}_{x' \rightarrow x} \frac{1}{4 \pi^2} \frac{1}
      {{(\sigma^\lambda \sigma_\lambda)}} \; 
      \frac{\chi}{2},
   \nonumber \\
   p_{quadratic} &  = & 
      {\lim}_{x' \rightarrow x} \frac{1}{4 \pi^2} \frac{1}
      {{(\sigma^\lambda \sigma_\lambda)}} \;
      \frac{\chi}{2} \; \frac{1}{3},
   \nonumber \\
   \epsilon_{logarithmic} & = & -
      {\lim}_{x' \rightarrow x} \frac{1}{4 \pi^2} \; 
      \frac{\chi^4}{8}
      \left( \gamma + \frac{1}{2} \ln \mid\frac{\chi \sigma}{2} \mid
      \right),
   \nonumber \\
   p_{logarithmic} & = & 
      {\lim}_{x' \rightarrow x} \frac{1}{4 \pi^2} \; 
      \frac{\chi^4}{8}
      \left( \gamma + \frac{1}{2} \ln \mid\frac{\chi \sigma}{2} \mid
      \right)
\>.
\label{eq:e-pcovariant}
\end{eqnarray}
Thus it follows
\begin{eqnarray}
   \epsilon_{quadratic} & = & 3 \; p_{quadratic},
   \nonumber \\
   \epsilon_{quartic} & = & - 3 \; p_{quartic},
   \nonumber \\
   \epsilon_{logarithmic} & = & - \; p_{logarithmic}
\>.
\label{eq:e-pnoncovariant}
\end{eqnarray}
We can now compare these general results for spatial point-splitting
with Eqs.~(\ref{eq:edivergent}) and (\ref{eq:pdivergent}) by making
use of the natural identification $\sigma^\lambda \sigma_\lambda
\propto \Lambda^{-2}$. It is easy to see that the quartic, quadratic
and logarithmic terms of these expressions fulfill the requirements of
Eq.~(\ref{eq:e-pnoncovariant}), as we wanted to show.

%


\end{document}